\documentclass[aps,pra,tightenlines,twocolumn,superscriptaddress,notitlepage,superscriptaddress,nofootinbib,shortbibliography]{revtex4-1}
\usepackage{graphicx}      
\usepackage{subfigure}
\usepackage[caption=false]{subfig}
\usepackage{epstopdf}             
\usepackage{amssymb}
\usepackage{mathtools}   
\usepackage{soul,xcolor} 
\usepackage{stackengine} 
\usepackage{placeins}

\usepackage{multirow}  
\usepackage{lipsum}
\usepackage{enumitem}     
\usepackage{xfrac}
\usepackage[normalem]{ulem}

\newcommand{\bra}[1]{\langle #1 |}

\newcommand{\ket}[1]{| #1 \rangle}

\newcommand{\SI}{Supplementary Material}  
\newcommand{\FigSI}[1]{S{#1}}
\newcommand{\SecSI}[1]{{#1}}

\newcommand{\CR}{'consistency region'{}}
\usepackage{ulem}  
\newcommand{\oalex}[1]{{\color{blue}{}}}

\definecolor{olive}{RGB}{107,142,35}

\newcommand{\oms}[1]{{\color{orange}{}}}

\newcommand{\RNum}[1]{\uppercase\expandafter{\romannumeral #1\relax}}

\begin{document}   
\title{Gaussian Process Regression for Absorption Spectra Analysis of Molecular Dimers}
\author{Farhad Taher-Ghahramani}
\email{farhadtq@pks.mpg.de}
\affiliation{Max Planck Institute for the Physics of Complex Systems, N\"{o}thnitzer Str 38, Dresden, Germany}

\author{Fulu Zheng}
\email{fzheng@uni-bremen.de}
\affiliation{Bremen Center for Computational Materials Science, University of Bremen, Am Fallturm 1, 28359 Bremen, Germany}

\author{Alexander Eisfeld}
\email{eisfeld@pks.mpg.de}
\affiliation{Max Planck Institute for the Physics of Complex Systems, N\"{o}thnitzer Str 38, Dresden, Germany}


\begin{abstract}
A common task  is the determination of system parameters from spectroscopy, where one compares the experimental spectrum with calculated spectra, that depend on the desired parameters.
Here we discuss an approach based on a machine learning technique, where the parameters for the numerical calculations are chosen from  Gaussian Process Regression (GPR). 
This approach does not only quickly converge to an optimal parameter set, but in addition provides information about the complete parameter space, which allows for example to identify extended parameter regions where numerical spectra are consistent with the experimental one.  
We consider as example dimers of organic molecules and aim at extracting in particular the interaction between the monomers, and their mutual orientation.
We find that indeed the GPR gives reliable results which are in agreement with  direct calculations of these parameters using  quantum chemical methods.

\end{abstract}

\maketitle    
\setstcolor{red}

\section{Introduction}

Because of their collective excitations, molecular aggregates, i.e., assemblies of long-range interacting chromophores, show remarkable optical properties and can exhibit efficient transport of electronic excitations (see e.g.,~\cite{saikin2013photonics,hestand2018expanded,Brixner2017Agg,Spano2020Agg}). 
To understand and design these properties a detailed knowledge of the arrangement of the molecules and their vibronic transitions is crucial. 
In this respect, absorption spectroscopy has been widely utilized to extract dynamical and geometrical information~\cite{abramavicius2009coherent,ginsberg2009two,schroter2015exciton,gao2018near,Ye2010MBO,Chen2016_MBO}.
Extracting the desired parameters from the spectra is often a demanding task.

Typically, there is a theoretical model for the aggregate, from which one can calculate the optical spectrum.
This model has various parameters (e.g., molecular positions, orientations of transition dipoles, vibrational frequencies).
One then varies these parameters and performs numerical calculations until a reasonable agreement between the calculated spectrum and the experimental one is obtained.

The experimental spectra are often noisy and the theoretical models rely on approximations.
Therefore, an exact agreement between theoretical and experimental spectra is in general not achievable. 
Instead, it is desirable to find the region(s) of parameter space that lead to a reasonable agreement between theory and experiment. 
To quantify the term 'reasonable agreement' one can introduce a measure for the distance between two spectra (e.g., the integrated absolute overlap or the integrated absolute difference).
Within an optimization perspective, we refer to this measure as cost function and the values of the measure as the cost.
Using this terminology, the above problem can be phrased as follows:
We are interested in {\it all} parameter sets that give a cost smaller than a certain value $C_{\rm ref}$. 
For cost values $C\lesssim C_{\rm ref}$ there is 'reasonable agreement' between the spectra. 
In the following, we refer to the cost as function over the parameter space as `cost landscape', and the region in the cost landscape which lies below $C_\mathrm{ref}$, as the \CR.

Typically, optimization methods such as direct statistical methods (e.g., sum rules~\cite{eisfeld2007simple}), classical optimization methods (e.g., differential evolution~\cite{pishchalnikov2019orange}) to more recent machine learning methods~(e.g., artificial neural networks~\cite{ghosh2019deep}), provide only one 'optimal' point (sometimes with a crude error estimation), but do not provide information on the cost landscape. 
A brute-force scan of cost landscape is often computationally very expensive, in particular, when individual calculations are already expensive and the model has several parameters.
One optimization method that can provide the estimates of the {\it full} cost landscape is based on Gaussian Process Regression (GPR)~\cite{rasmussen2006gaussian}. A GPR optimizer performs Bayesian updates to a Gaussian prior model by fitting known data. In physics context, GPR was employed to examine, among others, variational quantum simulators~\cite{kokail2019self}, ultra-cold-atom experiments~\cite{wigley2016fast}, quantum control algorithms~\cite{sauvage2020optimal}, inter-atomic potentials~\cite{bartok2015g}, to predict quantum dynamical cost landscapes~\cite{dalgaard2021predicting}  and to construct cost landscapes for the laser parameters driven Rydberg aggregates used for quantum simulation~\cite{bentley2018gaussian}.

In this paper, we outline an efficient method based on GPR for extracting physical parameters, and corresponding cost landscapes, of molecular aggregates from spectra. 
We exemplify the procedure for the case of molecular dimers.
Here we focus on recent experiments where spectra with well-defined arrangements have been recorded. 
This allows us to compare our parameters, extracted from the spectra, with those obtained by using the known molecular arrangement.
We focus on the experiments of Ref.~\cite{margulies2016enabling}, where monomer and dimer spectra have been recorded with a controlled change of the arrangement of the dimer. 
For these experiments we will outline the usage of GPR and provide a detailed discussion of the obtained parameters.
In the supplement we present the GPR results of two other experiments. 
The first one is the experiment on a cyanine monomer and its dimer in water~\cite{baraldi2002dimerization}.
The second one is an experiment on  a perylene bisimide compound in $\rm{CHCl_3/MCH}$ solvent~\cite{son2014spectroscopic}.

The structure of the paper is as follows:
In section \ref{sec:Model} we introduce the theoretical model of the monomer and the dimer.
In section \ref{sec:spectra} we provide our method of calculating the spectra and introduce additional relevant parameters.
In section \ref{sec:SA} we then describe our GPR based procedure to extract the relevant system parameters from the spectra.
In section \ref{sec:extraction} we exemplify our procedure for the case of experimentally measured monomer and dimer spectra.
We conclude in section \ref{sec:conclusions}.
Additional details are provided in the \SI. There also more experimental dimer spectra are analysed using GPR.

\section{Model for Monomer and Dimer \label{sec:Model}}

We consider dimers composed of identical monomers. 
Each constituent monomer is described by an electronic two-level system coupled to one vibrational mode. 
The two monomers interact via transition dipole-dipole force. 
Below we specify the respective Hamiltonians for the monomer and the dimer.

\subsubsection{Monomer}

The spectra of the monomers considered are dominated by one vibrational progression. Denoting the electronic ground state by $g$ with energy $\epsilon_g$ and the relevant electronic excited state by $e$ with energy $\epsilon_e$, we write the Hamiltonian of the monomer as~\cite{mukamel1995principles}
\begin{equation}
\label{eq:Hmon}
H^\mathrm{ex}_\mathrm{mon}=\epsilon_e\ket{e}\bra{e}+\omega_\mathrm{vib}a^\dagger a+\sqrt{S}\,\omega_\mathrm{vib}\ket{e}\bra{e}  (a^\mathrm{\dagger}+a), 
\end{equation}
\noindent with $\omega_\mathrm{vib}$ as the frequency of the vibration and $S$ as the Huang-Rhys factor.
With $a$ and $a^\dagger$ we denote the usual annihilation and creation operators of a harmonic oscillator. 
Initially, the monomer is prepared in the ground state $\ket{\psi(0)}=\ket{g}\ket{\chi_g}$, where $\ket{\chi_g}$ denotes the ground state of the vibrational mode $a$. 

\subsubsection{Dimer}

We describe the dimer using a widely employed model~\cite{mukamel1995principles,eisfeld2005vibronic,may2011charge,Polyutov2012Dimer,Kasha1965Dimer}. The dimer is described by the Hamiltonian
\begin{equation}  
H^\mathrm{ex}_{\rm{dim}}=H_{\rm{el}}+H_{\rm{vib}}+H_{\rm{el-vib}}.
\end{equation}
\noindent Here, the electronic Hamiltonian contains the single exciton states $\ket{1}=\ket{e}\ket{g}$ and $\ket{2}=\ket{g}\ket{e}$ as
\begin{equation}
H_{\rm{el}}=(\epsilon_e+\delta)\sum_{n=1}^2\ket{n}\bra{n}+V\big(\ket{1}\bra{2}+\ket{2}\bra{1}\big),
\end{equation}
\noindent where $\epsilon_e$ is the monomer excitation energy, $V$ is the (transition dipole-dipole) Coulomb interaction and $\delta$ is an overall energy shift due to the dimerization.

\noindent The vibrational Hamiltonian contains the vibrational modes of the individual monomers,
\begin{equation}
H_\mathrm{vib}=\sum_{n=1}^2\omega_\mathrm{vib}a_n^\dagger a_n.
\end{equation}
\noindent and the interaction between electronic excitations and vibrations is given by
\begin{equation}
H_\mathrm{el-vib}=\sum_{n=1}^2\sqrt{S}\omega_\mathrm{vib}\,\ket{n}\bra{n} (a_n^\dagger+a_n).
\end{equation}
\noindent The dimer is initially in the ground state $\ket{\psi(0)}=\ket{g}\ket{g}\ket{\chi_0}$, where $\ket{\chi_0}=\ket{\chi_g}\ket{\chi_g}$. 
\begin{figure}\centering
\includegraphics[scale=0.38]{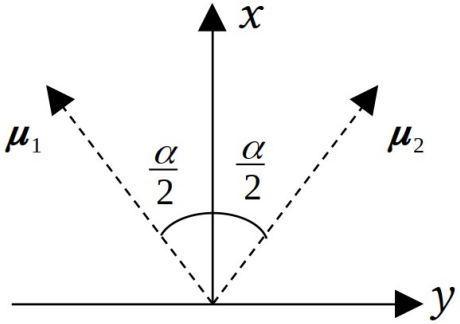}
\caption{\label{fig:geometry} The sketch of the dimer geometry. Two transition dipoles, $\boldsymbol{\mu}_1$ and $\boldsymbol{\mu}_2$ are located in the $x$-$y$ plane at an angle $\alpha$ with respect to each other.} 
\end{figure}

\section{Calculation of Spectra}\label{sec:spectra}
\subsection{Basic Equations}

The frequency-dependent absorption strength of an isotropically-oriented sample is determined as~\cite{may2011charge}
\begin{equation}
\label{eq:Abs}
A(\nu)=\mathrm{Re}\int_0^\infty dt~e^{i\nu t}M(t)W_{\sigma}(t),
\end{equation}
with the dipole correlation function
\begin{equation}
\label{eq:dipCorr}
M(t)=\bra{\psi(0)}\hat{\boldsymbol{\mu}}e^{-iH^\mathrm{ex} t}~\hat{\boldsymbol{\mu}}\ket{\psi(0)},
\end{equation}
where $W_{\sigma}(t)=e^{-\sfrac{\sigma^2t^2}{2}}$ is a Gaussian line-broadening function with width $1/\sigma$, $\hat{\boldsymbol{\mu}}$ is the total dipole operator, and $\ket{\psi(0)}$ is the initial state. The excited state Hamiltonian, $H^\mathrm{ex}$, is either that of the monomer or that of the dimer.

\subsubsection{Monomer}

The spectrum is calculated according to Eq.~(\ref{eq:dipCorr}) with
\begin{equation}
\label{eq:monM}
M_\mathrm{mon}(t)=\mu^2 \bra{e}\bra{\chi_g}e^{-i H_\mathrm{mon}^\mathrm{ex}t}\ket{\chi_g}\ket{e}. 
\end{equation}

\subsubsection{Dimer}

Here, the correlation function in Eq.~(\ref{eq:dipCorr}) becomes
\begin{equation}
\label{eq:dimM}
M_{\mathrm{dim}}(t)=\sum_{n,m=1}^2\boldsymbol{\mu}_n\cdot\boldsymbol{\mu}_m\mathcal{C}_{nm}(t), 
\end{equation}
with monomer transition dipole moments $\boldsymbol{\mu}_n\!\!=\!\bra{e}\hat{\boldsymbol{\mu}}_n\ket{g}$ and the correlation function
\begin{equation}
\label{eq:dimC}
\mathcal{C}_{nm}(t)=\bra{n}\bra{\chi_0}e^{-iH_{\mathrm{dim}}^\mathrm{ex}t}\ket{\chi_0}\ket{m}.
\end{equation}
\noindent We choose a coordinate system in which $\boldsymbol{\mu}_1$ and $\boldsymbol{\mu}_2$ lie in the $\mathrm{x}$-$\mathrm{y}$-plane where the $\mathrm{x}$-axis bisects the torsional angle $\alpha$ between the two moments (see Fig.~\ref{fig:geometry}).
Then we obtain $\boldsymbol{\mu}_n\cdot\boldsymbol{\mu}_n=\mu^2$ and $\boldsymbol{\mu}_m\cdot\boldsymbol{\mu}_{n\neq m}=\mu^2\cos\alpha$. 

\noindent If we define
\begin{equation}
\label{eq:phi}
\ket{\phi_n(t)}=\bra{\chi_0}e^{-iH_{\mathrm{dim}}^\mathrm{ex}t}\ket{\chi_0}\ket{n},
\end{equation}
one can express Eq.~(\ref{eq:dimM}) as~\cite{ritschel2015non} 
\begin{equation} 
\label{eq:Abs_Dimer}
M_\mathrm{dim}(t)=(1+\cos\alpha) M_{+}(t) + (1-\cos\alpha) M_{-}(t),  
\end{equation}
with
\begin{equation}
M_{\pm}(t)=\frac{1}{2}\mu^2\big(\bra{1}\pm\bra{2}\big)\big(\ket{\phi_1(t)}\pm\ket{\phi_2(t)}\big).
\end{equation}
Thus, the absorption spectrum can be calculated from a weighted sum of symmetric and anti-symmetric contributions. Note that also the sign and magnitude of $V$ depends on the angle $\alpha$. Thus, the absorption spectrum of dimer depends strongly on the angle $\alpha$. For $V=0$ (and thereby $\delta=0$), the dimer spectrum is reduced to that of monomer. We finally note that we normalize all spectra by area, i.e., by $\int A(\nu) \mathrm{d}\nu$.

\subsection{Numerical Propagation}

To calculate the correlation function appearing in Eq.~(\ref{eq:monM}) and Eq.~(\ref{eq:dimC}) we employ the Hierarchy of Pure States (HOPS) method \cite{ritschel2015non,suess2014hierarchy,zhang2016non}.  
The interaction of the electronic excitation and the vibrational mode is then described by an environmental correlation function
\begin{equation}
\label{expform}    
C(\tau)=S\omega_{\rm{vib}}^{2}~e^{-(\gamma+i\omega_{\rm{vib}})\tau}.
\end{equation}
 where beside the coupling strength $S$ and the vibrational frequency $\omega_\mathrm{vib}$, both defined in Eq.~(\ref{eq:Hmon}), we introduced a (small) phenomenological damping constant $\gamma$, which is also convenient for the numerical propagation. 
 Details of the HOPS method are given in section \RNum{1} of the \SI.

\begin{figure}\centering
\includegraphics[scale=0.5]{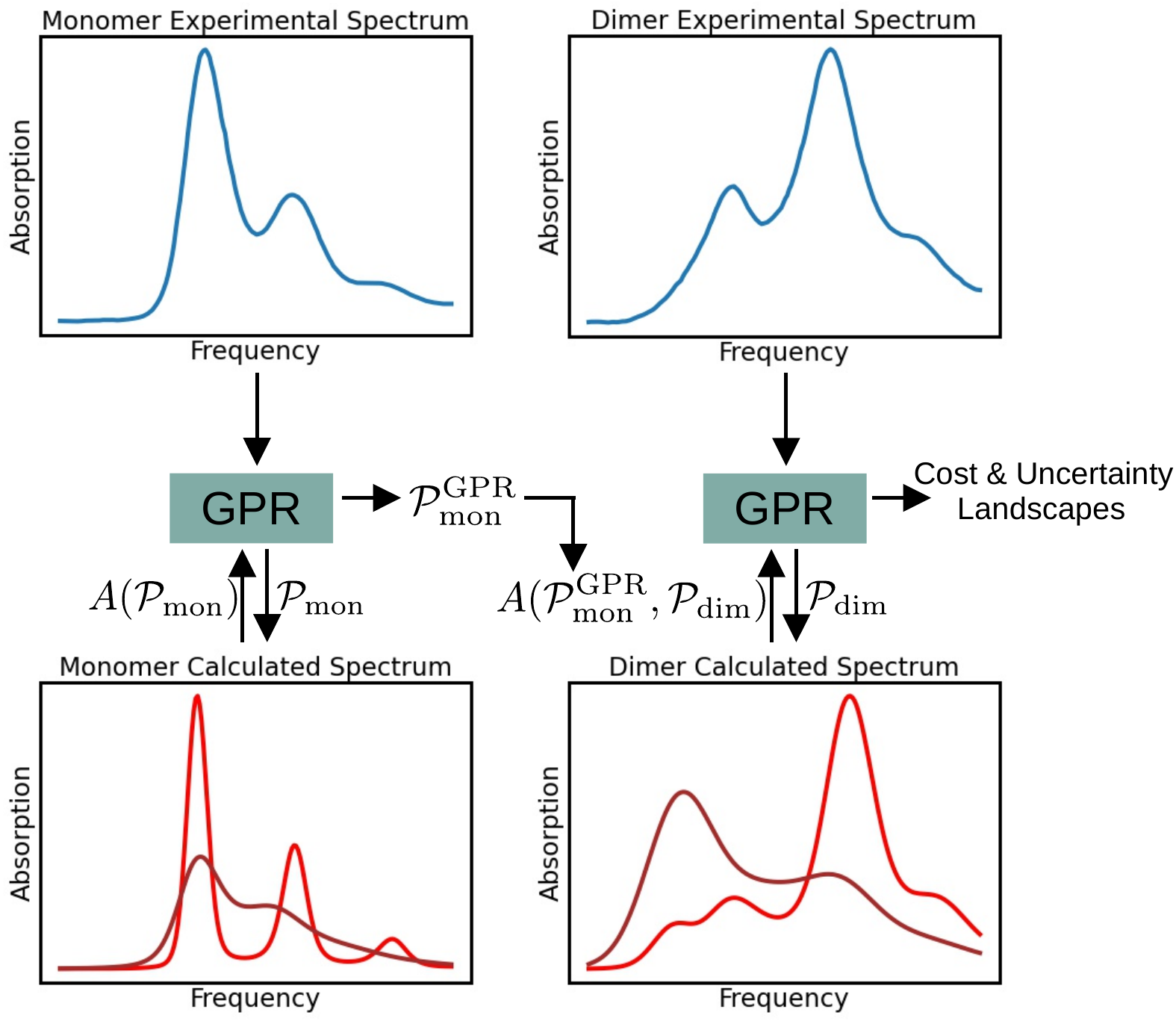}
\caption{\label{fig:sketch} Sketch of our procedure.
First, for the monomer parameter sets $\mathcal{P}_{\mathrm{mon}}$ are selected by the GPR method to explore the cost landscape and to find a spectra that agree well with the experimental one.
The predicted parameter set $\mathcal{P}_{\mathrm{mon}}^{\mathrm{GPR}}$ that agrees best is then used for the dimer calculations.
Also for the additional parameters of the dimer
 $\mathcal{P}_{\mathrm{dim}}$ 
we use GPR to find spectra that agree well with the experimental one.
The parameters  $\mathcal{P}_{\mathrm{dim}}$ are chosen such that they enable the construction of the complete cost landscape, which is important to find the complete \CR{}  of the parameters.
} 
\end{figure}
\section{Analysis of Spectra using Gaussian Process Regression}
\label{sec:SA}

As described above, each calculated absorption spectrum depends on a set of parameters, which we denote as $\mathcal{P}_\mathrm{mon}$ and $\mathcal{P}_\mathrm{dim}$ for the monomer and the dimer, respectively. Specifically, the monomer spectra depend on $\mathcal{P}_{\mathrm{mon}}=(\epsilon_e,\omega_{\rm{vib}},S,\gamma,\sigma_m)$ and the dimer spectra depends on $\mathcal{P}_{\mathrm{mon}}$ and $\mathcal{P}_{\mathrm{dim}}=(V,\delta,\alpha,\sigma_d)$. 
Here, $\sigma_\mathrm{m}$ and $\sigma_\mathrm{d}$, are the widths entering  the broadening function Eq.~(\ref{eq:Abs}) for the monomer and dimer, respectively.
The two parameter sets are summarized in Tab.~\ref{tab:params}.
\begin{table}\centering
\begin{tabular}{ccc}
\hline
Parameter Set & Quantity & Symbol \\
\hline
\multirow{6}{*}{$\mathcal{P}_{\mathrm{mon}}$} & Electronic Energy& $\epsilon_e$\\
& Vibrational Energy & $\omega_{\rm{vib}}$\\
& Vibronic Coupling & S\\
& Vibrational Damping& $\gamma$\\
& Broadening Width & $\sigma_m$\\
\hline
\hline
\multirow{4}{*}{$\mathcal{P}_{\mathrm{dim}}$} & Interaction Energy & V\\ 
& Interaction Energy Shift & $\delta$  \\  
& Torsional Angle&  $\alpha$ \\  
& Broadening Width & $\sigma_d$\\
\hline
\end{tabular}
\vspace{0.2cm}
\caption{\label{tab:params}Parameter sets for the monomer, $\mathcal{P}_{\mathrm{mon}}$, and the dimer, $\mathcal{P}_{\mathrm{dim}}$. }
\end{table}

Our main goal is to determine the parameters $V$ and $\alpha$ of the dimer. As stated above we are not so much interested in finding the optimal solution, but rather to explore the region of parameter space that gives reasonable agreement with the experimental spectra. To this end we use a procedure based on GPR~\cite{rasmussen2006gaussian}.

\subsection{Procedure}

The procedure consists of two successive GPR calculations. First we employ GPR to extract the monomer parameter set, $\mathcal{P}_{\mathrm{mon}}$, from the monomer experimental spectra, then introduce the optimized $\mathcal{P}_{\mathrm{mon}}$ in the dimer model and employ another GPR to extract the dimer parameter set, $\mathcal{P}_{\mathrm{dim}}$, from the corresponding dimer experimental spectra (see Fig.~\ref{fig:sketch}).

\subsection{Gaussian Process Regression}
\label{subsec:GPR}

 GPR fits known points in the cost landscape (i.e. parameter sets with known costs) to a Gaussian distribution. From the fitted distribution a Gaussian distribution, with the associated mean cost and standard deviation, is predicted for the cost at any given `test' $\mathcal{P}$. The covariance of predicted costs over parameter space depends on length-scales for characteristic variation in the cost as a function of each parameter. These length-scales are optimized to give the best fit of the landscape.

The numerical routine we use is the GPR optimizer of the M-LOOP package~\cite{wigley2016fast,bworld}; for further details see section \SecSI{\RNum{2}} of the \SI.

\indent The parameter sets $\mathcal{P}$ used for testing can be suggested by a so-called `acquisition function' according to different strategies: to reduce the predicted cost, or to reduce the predicted uncertainty, or a balance of these optimization-exploration criteria. Here we employ the balanced strategy by using an acquisition function of the form
\begin{equation}
\label{eq:aqfun}
f(\mathcal{P}, n_\mathrm{sc})=\mathrm{Cost^{\mathrm{GPR}}(\mathcal{P})}- n_{\mathrm{sc}}\cdot\mathrm{Uncer^{\mathrm{GPR}}(\mathcal{P})}
\end{equation}
where $\mathrm{Cost^{\mathrm{GPR}}(\mathcal{P})}$ and $\mathrm{Uncer^{\mathrm{GPR}}(\mathcal{P})}$ are the GPR-predicted cost and uncertainty (the standard deviation of the predictive distribution at query points.).
The scaling parameter $n_{\mathrm{sc}}$ is used to balance the two terms.
We typically adjust it after a few runs when one knows roughly the relevant magnitudes of the cost landscape and the magnitude of the corresponding uncertainties.
After each fit of the GPR, the acquisition function, Eq.~(\ref{eq:aqfun}), is minimized for four different values of $n_\mathrm{sc}$, resulting in four parameter sets used in the following calculations.
We choose always one of the four values of $n_\mathrm{sc}$ to be zero, which corresponds  to finding the minimal cost.
The other three values are chosen appropriately to sample larger parts of the surface, where one does not have much knowledge yet.

To evaluate how well the GPR algorithm works, we have calculated spectra and the corresponding costs for a large number of $N_\mathrm{tot}$ suitably chosen parameter sets.
Based on these we calculate the absolute difference of the GPR cost, $\rm{Cost}^{\rm{GPR}}$ and the exact one, $\rm{Cost}^{\rm{exact}}$, and average over parameter sets
\begin{equation}
\label{Eq:cost_conv}
\mathbb{E}^{\rm{C}} = \frac{1}{N_{\rm{tot}}}\sum_{j=1}^{N_\mathrm{tot}}
\big|\rm{Cost}^{\rm{exact}}(\mathcal{P}_j)-\rm{Cost}^{\rm{GPR}}(\mathcal{P}_j)\big|
\end{equation}
Beside this measure, that requires the knowledge of the exact cost values for many parameter sets, we also consider an intrinsinc measure based on the GPR uncertainty $\mathrm{Uncer}^{\rm{GPR}}$.
As in Eq.~(\ref{Eq:cost_conv}) we average over parameter points.  
\begin{equation}
\label{Eq:uncer_conv}
\Delta^{\rm{GPR}} = \frac{1}{N_{\rm{tot}}}\sum_{j=1}^{N_\mathrm{tot}}\rm{Uncer}^{\rm{GPR}}(\mathcal{P}_j)
\end{equation}
In the following we are interested in the landscapes over the complete parameter space over which the GPR is allowed to choose parameters. For each parameter we chose $25$ points, so we have $N_{\rm{tot}}=25^5$ for the monomer and $N_{\rm{tot}}=25^4$ for the dimer. 

\subsection{Distance Measure between Spectra}

To quantify how well the spectra are predicted, we define a distance measure as the difference between the calculated spectrum, $A^{\mathrm{cal}}$, evaluated at a given test $\mathcal{P}$, and the experimental one, $A^{\mathrm{exp}}$ as
\begin{equation}
\label{cost}
\mathrm{Cost}(\mathcal{P})=\int d\nu\big[A^{\mathrm{exp}}(\nu)-A^{\mathrm{cal}}(\nu;\mathcal{P})\big].
\end{equation}
We use this measure as the cost function of the GPR optimizer. Note that since the area of spectra is normalized to 1, the maximum value of the cost is 2.  

\section{Example: TDI monomer and TDI dimers with different arrangements}
\label{sec:extraction}

Now we employ the procedure outlined in the section~\ref{sec:SA} to extract the parameters and the corresponding cost landscapes from experimental spectra. 
Here we focus on the experiment of Ref.~\cite{margulies2016enabling}. More examples are presented in the \SI.

\begin{figure*}
  \centering
  \includegraphics[width=18cm]{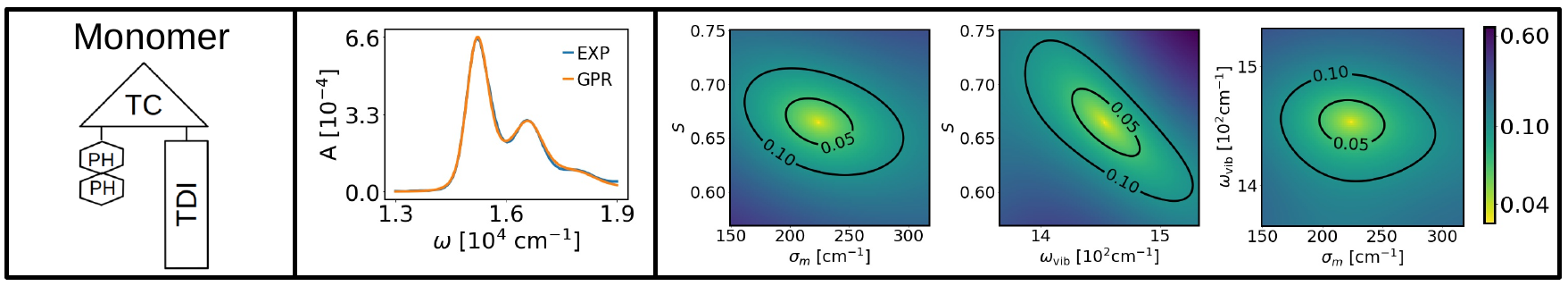}
  \caption
    {\label{fig:monomer} GPR predictions for the monomer of the experiment of Ref.~\cite{margulies2016enabling}.
    Left: sketch of the molecular arrangement. 
    Middle: best GPR spectrum (GPR) compared to the experimental one (EXP). 
    The parameters are given in Tab.~\ref{tab:values_mon}.
    Right: three cuts through the GPR cost landscape. 
    All cuts have as common point the parameter set of the best prediction given in Tab.~\ref{tab:values_mon} and are restricted to parameters $S,\omega_{\rm{vib}},\sigma_m$ around the relevant region around the optimal point. 
    }
\end{figure*}

In the experiment of Ref.~\cite{margulies2016enabling}, monomeric terrylene-3,4:11,12-bis(dicarboximide) (TDI) and three dimers  with different molecular arrangements  are examined. 
In the dimers, the two TDI molecules are bridged by a triptycene spacer and longitudinally shifted by inserting phenyl spacers on the bridge. 
The experimental spectra along with the corresponding molecular arrangements are presented in the left panel of Fig.~\ref{fig:monomer} and Fig.~\ref{fig:dimer}.
Details of the dimer structures are given in section \SecSI{\RNum{4}} of the \SI.

\subsection{Monomer}

First, we use GPR to extract the monomer parameter set $\mathcal{P}_{\mathrm{mon}}=(\varepsilon,S,\omega_{\rm{vib}},\gamma,\sigma_m)$ from the monomer experimental spectrum. 
We allow the GPR to search over a relatively large parameter range: $\varepsilon/(2\pi) \in[2400,2600]$,  $S\in [0.1,1]$, $\omega_{\rm{vib}}/(2\pi)\in [1,250]$, $\gamma \in [1,50]$, and $\sigma_m \in [100,1000]$. 
Here for the four scaling hyperparameters, we choose $n_{\mathrm{sc}}=0$, $1$, $2$, $3$.

\begin{table}[b]
\centering
\setlength{\tabcolsep}{6pt}
\begin{tabular}{ccccc}
\hline
$\varepsilon_e$($\rm{cm^{-1}}$) & $\omega_{\rm{vib}}$($\rm{cm^{-1}}$) & S & $\gamma$($\rm{cm^{-1}}$) & $\sigma_m(\rm{cm^{-1}})$ \\ 
\hline
16120 & 1450  & 0.67 & 37 & 223\\ \end{tabular}\caption{\label{tab:values_mon}The optimal parameters for the monomer found by GPR after evaluating 1000 parameter sets.
}
\end{table}

In Fig.~\ref{fig:monomer}, we summarized the GPR results. 
In the middle panel, we present the optimal spectrum found by the GPR together with the experimental one  after evaluation of 1000 parameter sets (the corresponding parameters are provided in Tab.~\ref{tab:values_mon}). 
One can see that the agreement between the GPR and the experimental spectrum is quite good. To get a feeling about the parameter regions in which one would get also reasonable agreement with experiment, in the right panel we show GPR cost landscapes. These plots show two-dimensional cross sections where the costs  are predicted for scans of two parameters at fixed GPR optimal values for the other parameters.
In section~\SecSI{III} of the \SI~we provide a comparison with exact cross sections where we calculate for each scanned parameter set the respective spectrum and calculate the corresponding cost with respect to the experimental spectrum.

\begin{figure}
  \centering
  \includegraphics[width=8cm]{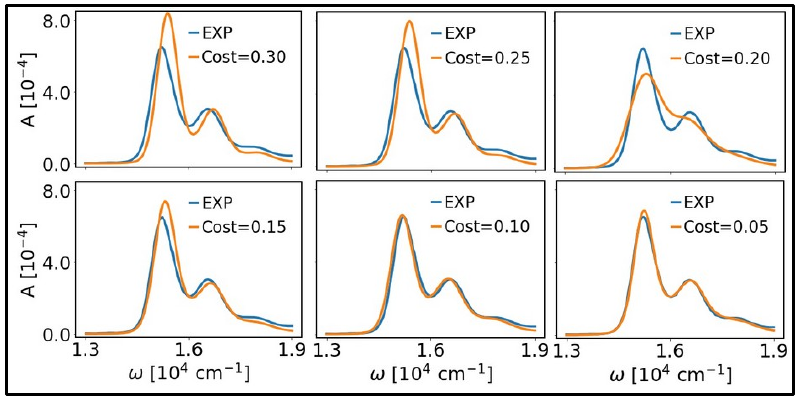}
  \caption
    {\label{fig:check_monomer} Examples of cost values for various calculated monomer spectra.}.
\end{figure}

To obtain a feeling of the cost values appearing in Fig.~\ref{fig:monomer}, and to specify the \CR , we show in Fig.~\ref{fig:check_monomer} examples of calculated spectra for various parameter sets together with their cost values.
We see that cost values $\lesssim 0.1$ lead to a reasonable agreement between two spectra.
We can now analyse the cost landscapes of Fig.~\ref{fig:monomer}.
For parameter sets with cost smaller then 0.1, i.e. inside the \CR , we have typically reasonable agreement with experiment.
That means that we have roughly a 5~\% uncertainty in the predicted values.
We see that the cost landscape is rather smooth, as expected for the monomer spectra.
From the cut $S$ versus $\omega_\mathrm{vib}$ we see, for example, that the cost increases much faster along the diagonal than along the off-diagonal.
For all cuts we see that our best prediction is located nicely in the center of the region with cost smaller than 0.05.
For the following dimer calculations we use the values of this spectrum, which are given in Tab.~\ref{tab:values_mon}.
We note that all parameter sets with cost smaller than 0.05 give very close agreement with the experimental spectrum. 
Therefore, they all could be used in the subsequent dimer calculations.
From calculations for different monomer parameters one could for example learn something about the dependence of the dimer parameters on the monomer ones.
We comment on this point further in the conclusions. 

 We now discuss the quality and convergence of the cost  landscape.
To judge the quality of the cost landscapes we compared to the exact cost landscape obtained by sampling the full four-dimensional parameter space on a fine grid.
We then calculate the mean of the absolute difference between the exact cost values and the estimated ones.
For the three cross-sections shown in Fig.~\ref{fig:monomer}, we present the exact cross section and the corresponding plots of the difference in Fig.~\FigSI{1} and Fig.~\FigSI{2} of the \SI, respectively.
We see that in the relevant \CR~of the parameter space the difference is very small (mostly smaller than 0.05), which demonstrates that the predicted cost-surfaces are highly accurate.
To quantify the global accuracy we have calculated the average of the absolute difference over the {\it complete} parameter grid (which is larger than the ranges shown in Fig.~\ref{fig:monomer}.
We find a value of around 0.2, which is much larger than the accuracy in the interesting region where one has small cost.
We also investigated this  global accuracy as a function of the GPR runs (see Fig.~\FigSI{4} of the \SI). 
We find that it fluctuates between values of 0.2 and 0.06, with no clear trend towards smaller values with increasing number of runs. 
We come back to this point below. 
Within the GPR approach one can estimate the uncertainty of the cost-landscape.
We find that for the same cuts as in Fig.~\ref{fig:monomer} these intrinsic estimates are roughly an order of magnitude smaller than the actual errors; (see Fig.~\FigSI{3} of the \SI).
This indicates that one should not rely too strongly on the absolute numbers of the intrinsic uncertainty of the GPR.
However, for the complete parameter range we find that the average estimated uncertainty agrees well with the one obtained from the difference of the estimated  and the actual cost.
We further note that our choices of the parameters $n_\mathrm{sc}$ in Eq.~(\ref{eq:aqfun}) actually leads only to little exploration of the parameter space, since the chosen values $n_\mathrm{sc}=0,\,1,\,2,\,4$ are so small that the term $n_\mathrm{sc}\cdot \mathrm{Uncer}^\mathrm{GPR}$ is small compared to the cost.
This might be the reason why the global difference between the estimated cost and the GPR cost is not decreasing with the number of runs.
For the dimers, where we use larger values of $n_\mathrm{sc}$ we observe such a decrease.

\subsection{Dimer}
We now come to the more interesting part; the prediction of the dimer parameters, in particular the angle $\alpha$ and the interaction strength $V$.
We use the monomer optimal parameter set $\mathcal{P}_{\mathrm{mon}}$, presented in Tab.~\ref{tab:values_mon}, to extract the dimer parameter set, $\mathcal{P}_{\mathrm{dim}}=(V,\delta,\alpha,\sigma_\mathrm{d})$ from the corresponding dimer experimental spectrum. 
The GPR optimizer searches over the parameter range $V/(2\pi) \in [1,250]$, $\sigma_d\in [100,1000]$, $\delta\in [-300,300]$ and $\alpha\in [0.1,180]$. 
For dimer 0 and 1 we used $n_\mathrm{sc}= 0,\ 20,\ 40,\ 60$. 
For dimer 2 we used $n_\mathrm{sc}= 0,\ 15,\ 30,\ 45$.

\begin{figure*}
  \centering
  \includegraphics[width=18cm]{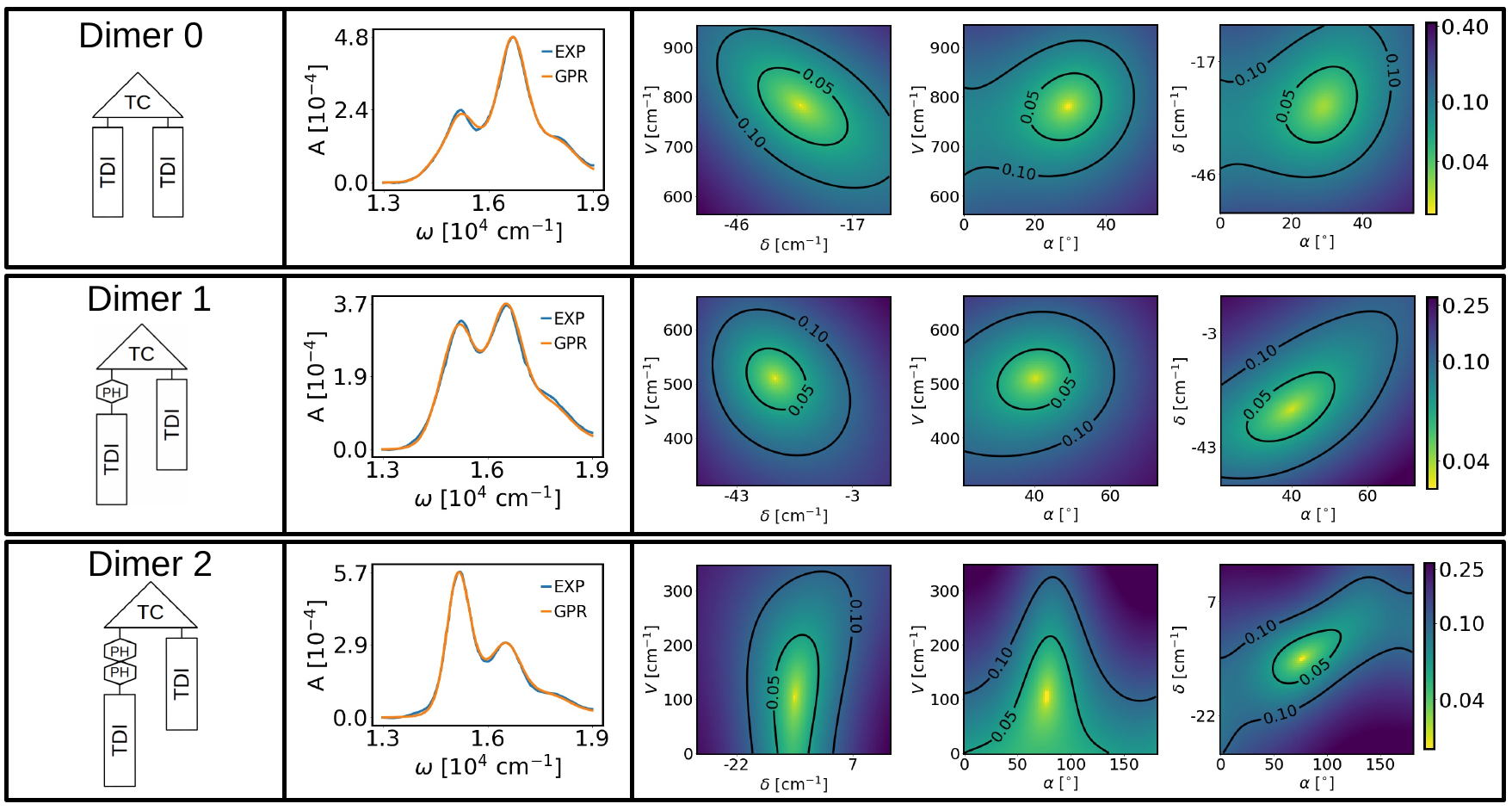}
  \caption
    {\label{fig:dimer} Same as Fig.~\ref{fig:monomer}, but for the three different dimers of the experiment of Ref.~\cite{margulies2016enabling}.
    To have a feeling for the cost values, in Fig.~\ref{fig:check_dimer} we present calculated spectra with their respective costs.
    The basic information about the parameters that can be extracted from the landscapes is provided in Tab.~\ref{tab:values_dimer}.}
\end{figure*}

\begin{table*}
\centering
\setlength{\tabcolsep}{5pt}
\begin{tabular}{c||cc||cc|cc|c|c}
\hline
Dimer& 
$V^\mathrm{DFT}$($\rm{cm^{-1}}$) &
$\alpha^\mathrm{DFT}\, (^\circ)$ &
$V\,(\rm{cm^{-1}}$) &
$[V_\mathrm{min},V_\mathrm{max}]$&
$\alpha$($^{\circ}$) &
$[\alpha_\mathrm{min},\alpha_\mathrm{max}]$ &
$\delta$($\rm{cm^{-1}}$) &  $\sigma_d(\rm{cm^{-1}})$\\
\hline
0 & 746 & 9 & 755 &[650,\ 850] & 28& [0,\, 50] &$-28\pm 20$ &  286 \\ 
1 & 489 & 12 & 507 & [400,\,600] & 40&[10,\, 60] &$-29 \pm 20$&   316 \\
2 & 62 & 11 & 111  &[0,\,300]  &70 &[0,\, 180] &$-7 \pm 10$&  260\\ 
\end{tabular}
\caption{\label{tab:values_dimer}The  parameters for the three dimers of the experiment of Ref.~\cite{margulies2016enabling}. With $V^\mathrm{DFT}$ and $\alpha^\mathrm{DFT}$ we denote the values obtained from the molecular structure and quantum chemical calculations as described in the main text. 
All other parameters are extracted from the GPR.
Here, $V$, $\alpha$, $\delta$, and $\sigma_\mathrm{d}$ denote the best values found by the GPR. 
The uncertainties are rough estimates based on the cost landscapes of Fig.~\ref{fig:dimer}. }
\end{table*}

In Fig.~\ref{fig:dimer} we summarize our findings.
As for the monomer we show in the middle box for each dimer the predicted spectrum with the lowest cost (the corresponding parameters are provided in Tab.~\ref{tab:values_dimer}).
The right box shows cuts through the cost  landscape.

\begin{figure*}
  \centering
  \includegraphics[width=14cm]{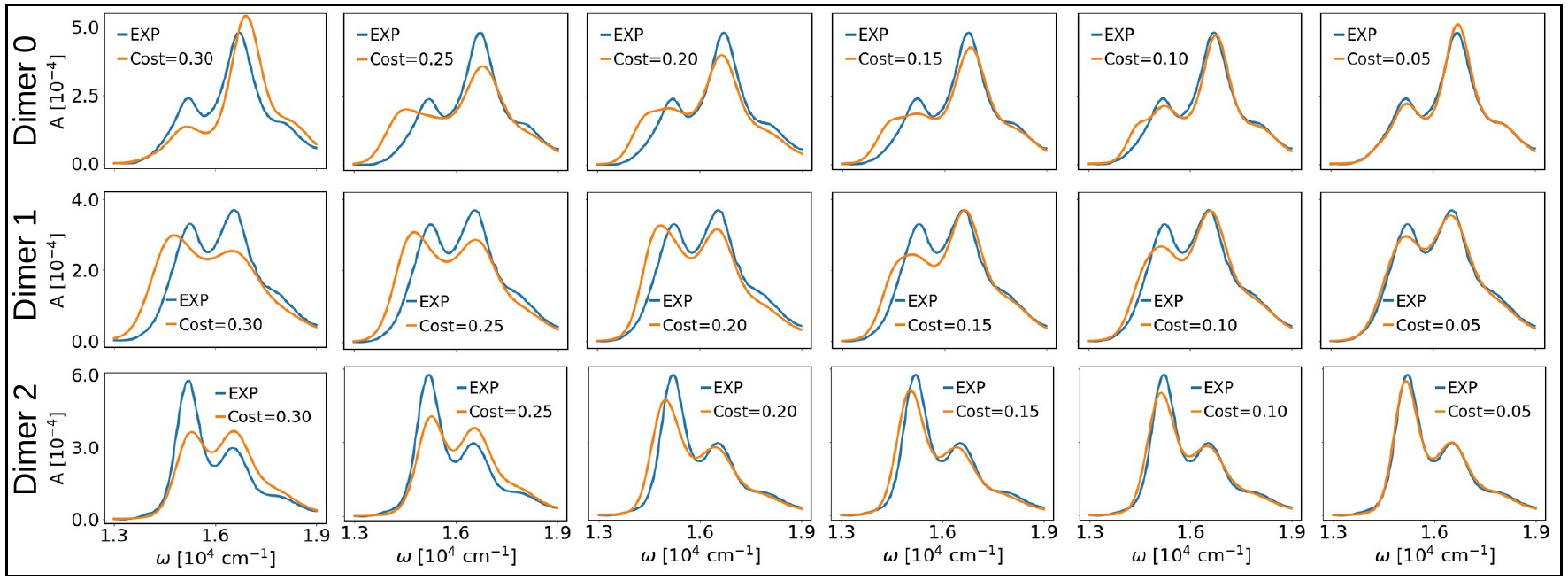}
  \caption
    {\label{fig:check_dimer}Examples of calculated dimer spectra (orange lines) with their respective cost. For each dimer the experimental spectrum is shown as the blue solid line. }
\end{figure*}

We now focus first on the best spectra and the corresponding parameters.
We see that in all three cases there is very good agreement between experiment and calculated spectra.
Looking at the parameters of Tab.~\ref{tab:values_dimer} we see that the interaction between the monomers decreases from  $V\approx 750\,\mathrm{cm}^{-1}$ for dimer 0, over $V\approx 500 \,\mathrm{cm}^{-1}$ for dimer 1 to a small value $V\approx 110\,\mathrm{cm}^{-1}$ for dimer 2.
The angle $\alpha$ increases from $\alpha \approx 30^{\circ}$ (dimer 0), over $\alpha \approx 40^{\circ}$ (dimer 1), to $\alpha \approx 80^{\circ}$ (dimer 2).
The shift $\delta$ is in all three cases quite small and the broadening $\sigma_\mathrm{d}$ is similar to that of the monomer, but slightly larger.
We now consider the cost landscapes.
As for the monomer, to obtain a feeling for the cost values in Fig.~\ref{fig:check_dimer} we show for all three dimers calculated spectra with their respective cost.
We see that for cost smaller than roughly 0.1 the agreement between simulation and experiment is reasonable, for larger cost values the agreement quickly gets worse.
Therefor we choose a cost of 0.1 to define the \CR{}.
For cost values smaller than 0.05 we found it difficult to judge by eye which spectra is better representing the experimental one.
For easier interpretation of the cost landscapes, we therefore have plotted the curves with constant cost of 0.05 and 0.1. 

We see that there is for all three dimers an extended region of parameters for which the GPR predicts good agreement.
The cost landscapes of dimer 0 and dimer 1 look quite similar.
Around the minimum the cost grows roughly  spherically symmetric.
The interaction $V$ can vary by approximately 10\% while still being in the region where the calculated spectra should agree with the experimental ones.
The angle $\alpha$ has a slight asymmetry, and the acceptable region extends nearly to zero degree. 
For dimer 2 the cost landscape  looks quite different.
Around the minimum there is an elongated region of permissible values of $V$ from 0 to $\sim 300\, \mathrm{cm}^{-1}$. 
For $V$ smaller than $100\, \mathrm{cm}^{-1}$ all values of $\alpha$ are within an acceptable cost.
That means, that the predicted value of $\alpha$ from the minimum is meaningless.
Here one clearly sees the strength of the GPR landscapes over a naive optimization where one would have taken the value of 80 degrees seriously.
 To see if the predictions from the GPR are trustworthy, we have again performed full scans of the parameter space using a fine regular grid.
The cuts corresponding to the ones of Fig.~\ref{fig:dimer} are presented in Fig.~\FigSI{1} of the \SI.
We find that indeed the GPR cross-sections resemble very accurately the exact ones.

To check if our predictions make sense,  we have performed independent calculations based on the molecular structures provided in Ref.~\cite{margulies2016enabling}.
   We first optimized the molecular structure of a TDI monomer (for 100 steps) and then calculated the transition charges using the DFTB+ package~\cite{aradi2007dftb+,hourahine2020dftb+}.
    Using these transition charges and the molecular structures of the dimers, as given in Ref.~\cite{margulies2016enabling}, we then evaluated  the interaction $V$ using the {\it Transition Charges from Electrostatic Potentials} (TrEsp) method, which has been widely applied to accurately evaluate inter-molecular couplings~\cite{howard2004approaches,patwardhan2010theoretical,kistler2013benchmark,zheng2016TrEsp}\footnote{Because of the short distance between two monomers, the often used  point-dipole approximation is invalid to evaluate the interactions between the dimers in this work and gives unreasonable results.}.
    The angle $\alpha$ is also estimated directly from the given dimer structures (for 
    details see section~\SecSI{\RNum{4}} of the \SI.
    For the values of $V$ we find very close agreement to the ones from GPR optimum. 
    However, for $\alpha$ we always find an angle around 10 degrees.  
This highlights again the importance of the cost landscape, which prevents wrong conclusions drawn from the optimal values.
For the case of our dimers, we see in particular that all $V$ values from the TrEsp method and the estimated angles $\alpha$ from molecular structures are within the 'reliable' region of the GPR predicted cost surface.
     
Let us briefly comment on the peculiar behaviour of the cost landscape of dimer 2 with respect to $\alpha$. 
For dimer 2 essentially all values of $\alpha$ give spectra that agree very well with the experimental one. 
We first note that the experimental spectrum is nearly identical to the monomer one.
That implies a very small coupling $V$. 
This results in two nearly identical spectra for the plus and minus components of Eq.~(\ref{eq:Abs_Dimer}).
Therefore, the angle $\alpha$ that weights the two components becomes irrelevant.
   
As for the monomer, also for the dimer we calculated spectra on a fine grid of the complete parameter space and determined the corresponding costs.
For this exact cost landscape we present cuts in Fig.~\FigSI{1} of the \SI{} for the same parameters as in Fig.~\ref{fig:dimer}.
We see that they look quite similar, which is quantified by taking the point-wise absolute difference (shown in Fig.~\FigSI{2} of the \SI).
In the relevant  region the difference is in the order of 0.02 to 0.05. 
We emphasize that all features relevant for the extraction of the dimer parameters are well reproduced by the estimated cost landscapes of Fig.~\ref{fig:dimer}.
Furthermore we found that for the complete parameter space we have an average absolute difference between the actual and predicted cost that decreases exponentially with the number of runs (with fluctuations), indicating that we explore the complete parameter space (see Fig.~\FigSI{3} of the \SI). 
After only 100 runs we have already an average error bellow 0.1.
For 1000 runs the average error is smaller than 0.03, which indicates that we can predict the surface also quite well far away from the minimum of the cost.

Also for the dimer we analysed the uncertainty of the cost values predicted by the GPR.
For the cuts these uncertainties are shown in Fig.~\FigSI{4} of the \SI.
As for the monomer the predicted uncertainty for these cuts is roughly one order of magnitude smaller than the actual one.
For the complete landscape the average predicted uncertainty is comparable to the actual error.

\section{Conclusion}
\label{sec:conclusions}
We have investigated the use of GPR to extract molecular parameters from experimental spectra by comparing with numerical calculations where these parameters explicitly enter the model.
The advantages of GPR over other optimizers are, that GPR is essentially independent from the initial values and that there is also little dependence on the allowed ranges of the parameters.
Most importantly, GPR seeks not only an optimal solution, but also provides information about the whole parameter space.

We have found that in all cases considered, GPR converges rather quickly with the number of runs (i.e.~ the number of calculated spectra) to good results, both for the optimal solution and also for the complete cost-landscape, which provides information about all parameter sets, for which calculated and experimental spectra are in acceptable agreement.  
We found that already for around 100 evaluation we had useful predictions of the GPR.
Such a fast convergence is important, when individual calculations become expensive, for example when treating larger aggregates.
In the present study the calculation of a single spectrum takes only a few seconds, which allowed us to investigate the performance of  GPR when varying  hyperparameters  which determine the search strategy of the GPR.
We found that the results show no strong dependence on the  choice of these hyper-parameters, indicating the robustness of the method.

For all cases considered in the present work we have a rather smooth cost landscape, which facilitates the construction of the cost landscape.
 However, the GPR is also able to predict more complicated landscape as demonstrated e.g.~in Ref.~\cite{bentley2018gaussian}.
Even for the present smooth cost landscapes, the GPR is very efficient.
When one would sample the parameter space on a regular grid with only ten points per parameter, one would need $10^5$ calculations for the monomer and $10^4$ for the dimer, much more than the $10^3$ that we have used to predict the cost landscape (already for around 200 evaluated parameter sets we have a reasonable good representation of the exact cost landscape.)

In the present study we had four parameters to optimize for the dimer and five for the monomer. 
In principle a larger number of parameters is possible. 
However, we believe that GPR is most suitable for less than 10 parameters; an experience that we have made in our previous work on wavefunction reconstruction \cite{zheng2019excitonic}.
That is also the reason why we used a  sequential procedure where we first determined parameters of the monomer, which then entered the calculations of the dimer.
 We could have used also an approach where we simultaneously fitted the monomer and dimer with 8 parameters. 
However, we believe that the present sequential strategy is more economical, since convergence in a smaller parameter space is usually faster.
    
In our sequential approach we have used the optimal solution of the monomer parameters found by the GPR as input for the dimer calculations.
To get an even more refined estimate of the range of valid dimer parameters, we could have made additional optimizations using monomer parameters within the regions that also gave good agreement between the calculated and the measured monomer spectrum.    
We emphasize once more that it is of crucial importance that the complete cost landscape is available.
In our opinion it is more important than the optimal solution, because from the cost-landscape one can easily see which parameters give results for which the cost is below a certain threshold that gives good agreement with the experimental observation.

Beside the experiments of Ref.~\cite{margulies2016enabling}, which are discussed in Sec.~\ref{sec:extraction} we have investigated also other experimental spectra.
In section \SecSI{V} of the \SI{} we present results for a cyanine monomer and its dimer in water~\cite{baraldi2002dimerization} and a perylene bisimide compound in a $\rm{CHCl_3/MCH}$ solvent \cite{son2014spectroscopic}.
The general findings for these two cases are very similar to what has been shown in Sec.~\ref{sec:extraction}.
The monomers have in particular very similar vibrational frequencies and Huang-Rhys factors $S$.
The dimer spectrum of Ref.~\cite{baraldi2002dimerization} is very similar to that of dimer 0 in Fig.~\ref{fig:dimer} and the dimer spectrum of Ref.~\cite{son2014spectroscopic} is very similar to dimer 1 in Fig.~\ref{fig:dimer}.
Thus, as expected, we find similar interaction $V$ of $900\, \mathrm{cm}^{-1}$ and $500\, \mathrm{cm}^{-1}$ for the respective cases.
 We also find that the cuts through the cost-landscape show the same features as the ones in Fig.~\ref{fig:dimer} for dimer 0 and dimer 1.
In particular the parameters of the angle that give small cost extend asymmetrically to $\alpha=0$, with a predicted minimum around 25 degrees and 40 degrees for the two experiments, respectively.   
    
Although the theoretical model that we used gives very good agreement between experiment and calculations, there is plenty room for improvement. 
For example one could use static disorder to obtain a more realistic model of the line broadening or one could take more than one vibrational mode explicitly into account \cite{RoEiDv11_054907_}.
However, as we have seen, for the present purpose a very simple model is sufficient to reliably extract the desired parameters.    

 Finally, we emphasize that GPR can also be applied to various similar types of problems where one has a model with some parameters and one wants to extract these parameters from given experimental data.

\acknowledgements
AE acknowledges support from the DFG via a Heisenberg fellowship (Grant No EI 872/5-1). FZ acknowledges the support from DFG RTG-2247 Quantum Machanical Materials Modelling. 

%

\end{document}